\documentclass[letterpaper]{article} 
\usepackage{aaai24}  
\usepackage{times}  
\usepackage{helvet}  
\usepackage{courier}  
\usepackage[hyphens]{url}  
\usepackage{graphicx} 
\usepackage{natbib}  
\usepackage{caption} 
\usepackage{algorithm}
\usepackage[noend]{algpseudocode}

\usepackage{color}
\usepackage{booktabs}
\usepackage{amsmath,amssymb,amsfonts}
\usepackage{graphicx}
\usepackage{textcomp}
\usepackage{xcolor}
\usepackage{tikz}
\usepackage{textcomp}
\usepackage{fancyhdr}
\usepackage{booktabs}
\usepackage{tabularx, booktabs}
\usepackage{float}
\def\BibTeX{{\rm B\kern-.05em{\sc i\kern-.025em b}\kern-.08em
    T\kern-.1667em\lower.7ex\hbox{E}\kern-.125emX}}

\usepackage{newfloat}
\usepackage{listings}
\DeclareCaptionStyle{ruled}{labelfont=normalfont,labelsep=colon,strut=off} 
\floatstyle{ruled}
\newfloat{listing}{tb}{lst}{}
\floatname{listing}{Listing}
\title{Advancing DDoS Attack Detection: A Synergistic Approach Using Deep Residual Neural Networks and Synthetic Oversampling
}
\author{
    Ali Alfatemi\textsuperscript{\rm 1},
    Mohamed Rahouti\textsuperscript{\rm 1},
    Ruhul Amin\textsuperscript{\rm 1},
    Sarah ALJamal\textsuperscript{\rm 1},
    Kaiqi Xiong\textsuperscript{\rm 2},
    Yufeng Xin\textsuperscript{\rm 3}
}
\affiliations{
     \textsuperscript{\rm 1}Computer and Information Science Department, Fordham University, NY, USA\\
    \textbraceleft \textrm{aalfatemi, mrahouti, mamin17, saljamal} \textbraceright @\textrm{fordham.edu}\\
    \textsuperscript{\rm 2}Cyber Florida, University of South Florida, FL, USA\\
    \textrm{xiongk@usf.edu}\\
    \textsuperscript{\rm 3}RENCI, University of North Carolina at Chapel Hill, NC, USA\\
    \textrm{yxin@renci.org}
}

\usepackage{bibentry}

\begin{document}

\maketitle

\begin{abstract}
Distributed Denial of Service (DDoS) attacks pose a significant threat to the stability and reliability of online systems. Effective and early detection of such attacks is pivotal for safeguarding the integrity of networks. In this work, we introduce an enhanced approach for DDoS attack detection by leveraging the capabilities of Deep Residual Neural Networks (ResNets) coupled with synthetic oversampling techniques. Because of the inherent class imbalance in many cyber-security datasets, conventional methods often struggle with false negatives, misclassifying subtle DDoS patterns as benign. By applying the Synthetic Minority Over-sampling Technique (SMOTE) to the CICIDS dataset, we balance the representation of benign and malicious data points, enabling the model to better discern intricate patterns indicative of an attack. Our deep residual network, tailored for this specific task, further refines the detection process. Experimental results on a real-world dataset demonstrate that our approach achieves an accuracy of 99.98\%, significantly outperforming traditional methods. This work underscores the potential of combining advanced data augmentation techniques with deep learning models to bolster cyber-security defenses.
\end{abstract}

\section{Introduction}

The digital transformation has significantly transformed various aspects of our lives, yet it has also ushered in numerous challenges. Among these, Distributed Denial of Service (DDoS) attacks have emerged as one of the most pernicious, casting a shadow on the aspirations of a connected future \cite{rahouti2022sdn}. These attacks, characterized by an overwhelming surge of traffic to a target server or network, effectively render services unavailable to genuine users. Recent statistics illuminate the gravity of this situation: a staggering 7.9 million DDoS attacks were recorded in just the first half of 2023 \cite{netscout2023}. This surge not only incapacitates services but also erodes the Quality of Service (QoS) that legitimate users are accustomed to.

Amidst this backdrop, the increasing penetration of modern networks, including Vehicular and Internet of Things (IoT) networks, into our daily lives becomes evident. These systems, underpinned by distributed edge clouds, have been architected to efficiently cater to latency-sensitive services for a sprawling IoT domain \cite{madi2021nfv}. However, the anticipated proliferation of IoT devices, projected to soar to approximately 26.4 billion connections by 2026 \cite{ericsson2021}, augments the potential attack surface. The innate vulnerabilities of many IoT devices, often bereft of comprehensive security protocols, offer a ripe landscape for cyber adversaries \cite{cvitic2021novel, cvitic2021boosting}. These devices, when manipulated, can be marshaled to orchestrate sweeping DDoS attacks on globally distributed edge servers \cite{madi2021nfv, he2021game, wang2018data}, posing challenges not only to individual users but also to the broader digital ecosystem.

Over the years, various machine learning frameworks have been championed for DDoS attack detection. Traditional stalwarts such as K-Nearest Neighbor (KNN) \cite{vu2008ddos}, Support Vector Machine (SVM) \cite{cheng2009ddos}, Random Forest (RF) \cite{wang2017research}, and Naive Bayes (NB) \cite{fadlil2017review} have been extensively explored. These models, by virtue of their design, classify patterns in network traffic to discern anomalies. Nevertheless, their efficacy tends to wane when confronted with evolving and intricate DDoS attack paradigms.

In pursuit of more robust solutions, recent research endeavors have gravitated towards deep learning. The innate capability of deep learning architectures to understand and represent complex data relationships offers a promising avenue. Convolutional Neural Networks (CNNs), with their astute ability to parse raw data patterns, are being actively researched for DDoS detection \cite{hasan2018burst}. Deep Autoencoders, focusing on the creation of condensed data representations, hold potential in gleaning subtle, malicious traffic patterns \cite{krishnan2019varman}. Furthermore, Artificial Neural Networks (ANNs) \cite{ahanger2017effective, alzahrani2018detection} utilize nonlinear activations to delineate intricate relationships between varied data inputs, offering nuanced insights into potential threats.

The comparative prowess of deep learning methodologies over their traditional counterparts is underscored by recent empirical studies. For instance, a CNN model demonstrated an accuracy rate of 92\% on the NSL-KDD dataset, trouncing both SVM, which clocked 87\%, and KNN, which registered 80\% \cite{hasan2018burst}. Deep Autoencoders, in another study, achieved an accuracy of 99.6\% on the CIC-IDS2017 dataset, underscoring their potential for DDoS detection \cite{krishnan2019varman}. However, the cybersecurity landscape is ever-evolving, with attackers perpetually innovating their strategies. This necessitates a continuous research impetus to stay abreast and ensure robust defense mechanisms.

In the intricate tapestry of cybersecurity challenges, DDoS attack detection stands out as an area demanding innovative solutions. This paper embarks on a comprehensive journey to unravel a pioneering approach to DDoS attack detection. At the heart of our proposed methodology lies a harmonious fusion of the Synthetic Minority Over-sampling Technique (SMOTE) and the advanced capabilities of Deep Residual Neural Networks (ResNets). This blend is meticulously designed to craft a detection mechanism that encapsulates robustness, efficiency, and adaptability, ensuring resilience against the ever-evolving cyber adversarial tactics. Through the course of this paper, we strive to provide an in-depth examination of this approach, highlighting its transformative potential in the realm of DDoS attack detection.

The contributions of this paper are summarized as follows:
\begin{enumerate}
    \item Novel detection fusion: We present a unique integration of SMOTE with Deep ResNets, offering a fresh perspective to DDoS attack detection mechanisms in the literature.
    
    \item Adaptability: Our proposed model is not just a static solution; it's designed with the foresight to adapt to dynamic cyber threats, ensuring its relevance and efficacy in the face of evolving attack strategies.
    
    \item Comprehensive evaluation: Through rigorous experiments on benchmark datasets, we demonstrate the superiority of our approach over existing methods, substantiating our claims with empirical evidence.
    
    \item Practical implications: Beyond theoretical contributions, this paper underscores the real-world applicability of the proposed method, emphasizing its potential to bolster cybersecurity defenses in diverse operational environments.
\end{enumerate}

The subsequent sections of this article are methodically laid out to provide readers with a comprehensive understanding of the topic. Initially, \textbf{Related Work} Section offers an overview of the current state-of-the-art findings in the field. This is followed by \textbf{Research Background} Section \ref{sec:background}, which presents the theoretical background of the proposed approach. Next, \textbf{Methodology} Section delves into the implementation specifics of the proposed methodologies and techniques that have been employed for the research problem presented in this work. Moving forward, \textbf{Results and Discussion} Section presents the pivotal evaluation results, complemented with in-depth discussions to provide clarity and context. Last, \textbf{Conclusion} Section gives a concluding summary and encapsulates the essence of this work.

\section{Related Work} \label{sec:related}
DDoS attacks remain one of the most formidable challenges in network security \cite{rahouti2021synguard, owusu2023enhancing}. Traditional detection methods often struggle with the constantly evolving nature of these attacks, which are becoming more sophisticated and distributed \cite{zhao2023enhancing}. The rise of Deep Learning, particularly Deep ResNets, offers a promising avenue to tackle this issue \cite{ghillani2022deep}. ResNets, introduced by He et al. \cite{he2016deep}, have the unique capability of training very deep networks by introducing skip connections, which address the vanishing gradient problem, thereby enhancing performance \cite{shaikh2022real}.

One major challenge in training deep neural networks for DDoS detection is the imbalance in datasets \cite{zhu2022oversampling}. Real-world network traffic data often contains a significantly larger number of benign samples compared to malicious ones. This can result in models that are biased towards benign classifications. To counteract this data unbalanced issue, researchers have turned to synthetic oversampling techniques. These methods artificially generate new samples by mimicking the characteristics of the minority class, aiming to balance the class distribution \cite{wei2022improved}. Such techniques, when combined with ResNets, have been shown to significantly improve the accuracy and recall rates for detecting DDoS attacks \cite{hussain2020iot}.

Recent advancements in synthetic oversampling have considered the unique structures and patterns of network traffic data. Techniques like the adapted SMOTE for network traffic have been researched to better fit the characteristics of DDoS traffic \cite{xu2023data}. Furthermore, the integration of oversampling within the training pipeline of ResNets, rather than as a pre-processing step, has been a recent innovation \cite{dablain2022deepsmote}, allowing the network to learn more discriminative features.

In brief, the integration of synthetic oversampling techniques with Deep ResNets represents a cutting-edge approach to enhance DDoS attack detection. By addressing dataset imbalance directly within the deep learning framework, our proposed methodology holds promise in creating more robust and accurate DDoS detection models in the face of evolving threats.

\section{Research Background} \label{sec:background}
This section presents the theoretical background and abstraction of the proposed methodology.

Let $D = \{(x_1, y_1), (x_2, y_2), \dots , (x_n, y_n)\}$ be the dataset where
\begin{itemize}
    \item $x_i$ is the feature vector of the i-th sample.
    \item $y_i$ is the corresponding label where $y_i = 1$ denotes a DDoS attack and $y_i = 0$ denotes benign traffic.
\end{itemize}
\subsection{Synthetic Minority Over-sampling Technique (SMOTE)}
SMOTE generates synthetic samples in the feature space. For each minority sample $x$,
\begin{enumerate}
    \item choose one of the k nearest neighbors $x_{nn}$ from the minority class,
    \item compute the difference between the feature vectors $x$ and $x_{nn}$: $\text{diff} = x_{nn} - x$,
    \item multiply the difference by a random number between 0 and 1: $gap = \text{diff} \times \text{random}(0, 1)$, and
    \item add the gap to the original sample to create the synthetic sample: $x_{syn} = x + gap$.
\end{enumerate}
Mathematically, for each $x_i$ where $y_i = 1$:
\[ x_{syn} = x_i + \text{random}(0, 1) \times (x_{nn} - x_i) \]
\subsection{Deep Residual Network}
Given that ResNet structures leverage skip connections, the forward propagation for a basic block can be represented as:
\[ F(x) = \mathcal{F}(x, W_i) + x \]
where
\begin{itemize}
    \item $ F(x) $ is the output of the block.
    \item $ \mathcal{F}(x, W_i) $ represents the residual mapping (could be multiple convolutional layers).
    \item $ W_i $ are the weights of the block.
    \item $ x $ is the input to the block.
\end{itemize}
The overall ResNet model can be described in terms of two phases, training phase and detection phase.
\textbf{Training phase:} decomposed of two key steps as follows:
\begin{itemize}
    \item Apply SMOTE on $D$ to get a balanced dataset $D_{\text{balanced}}$.
    \item Train the ResNet model on $D_{\text{balanced}}$ to minimize the loss function $L$, which could be binary cross-entropy for a binary classification task:
    \[ L = -\frac{1}{n}\sum_{i=1}^{n}[y_i \log(\hat{y_i}) + (1-y_i) \log(1-\hat{y_i})] \],
    where $ \hat{y_i} $ is the model's prediction for the i-th sample.
\end{itemize}
\textbf{Detection phase:}
For a new traffic sample $x$, if the output of the network $ \hat{y} > \text{threshold} $, it is classified as a DDoS attack, otherwise benign.

\subsection{Attention-Augmented Residual Network}
To capture the interdependencies between different features in the network traffic, we use an attention mechanism. For a layer's output \( Z \):
\begin{align*}
A & = \text{Softmax}(W_a Z + b_a) \\
Z' & = A \odot Z
\end{align*}
where
\begin{itemize}
    \item \( W_a \) and \( b_a \) are the attention's weight and bias.
    \item \( A \) is the attention weights after softmax normalization.
    \item \( \odot \) represents element-wise multiplication.
    \item \( Z' \) is the attention-augmented output.
\end{itemize}
This mechanism can be integrated into several blocks of the ResNet, helping the network to focus more on significant features.

\subsubsection{Dual-Phase Training}
\paragraph{\textbf{Phase 1-- Pre-training on Original Data:}}
First, we train our attention-augmented ResNet on the original dataset \( D \). This objective function in this phase is the same binary cross-entropy:
\[ L_1 = -\frac{1}{n}\sum_{i=1}^{n}[y_i \log(\hat{y_i}) + (1-y_i) \log(1-\hat{y_i})] \]
\paragraph{\textbf{Phase 2-- Refinement with Augmented Data:}}
Post pre-training, the model is further trained on \( D_{\text{balanced}} \), which is the SMOTE-augmented dataset. However, we introduce a regularization term to the loss function, penalizing large deviations from the initial predictions (from Phase 1). Let \( \hat{y_i}^{(1)} \) be the prediction from Phase 1 and \( \hat{y_i}^{(2)} \) be the prediction in Phase 2 for sample \( x_i \):
\begin{align*}
L_2 = & -\frac{1}{n}\sum_{i=1}^{n}[y_i \log(\hat{y_i}^{(2)}) + (1-y_i) \log(1-\hat{y_i}^{(2)})] \\
      & + \lambda \sum_{i=1}^{n}(\hat{y_i}^{(1)} - \hat{y_i}^{(2)})^2, 
\end{align*}
where \( \lambda \) is a regularization hyperparameter.
The enhanced model consists of the attention-augmented ResNet trained with the dual-phase process described above. 

\begin{figure*}[h]
    \centering
    \includegraphics[width=\textwidth]{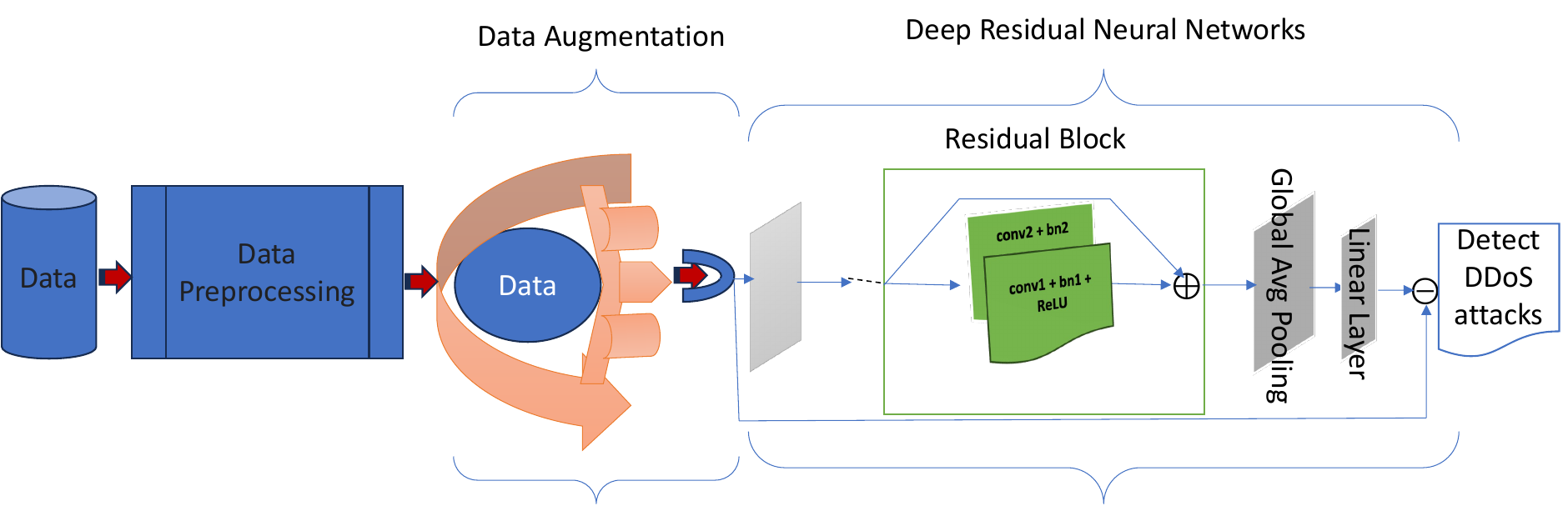}
    \caption{Schematic representation of the data processing and analysis pipeline for DDoS attack detection. The detection workflow starts with initial data preparation. This is succeeded by synthetic oversampling via SMOTE to rectify class disproportion. Subsequently, a deep residual neural network processes the equilibrated data to recognize complex patterns, culminating in the identification of DDoS intrusions.
    }
    \label{fig:Over}
\end{figure*}

The implemented methodology is depicted in Figure \ref{fig:Over}. This figure shows the schematic representation of the data processing and analysis pipeline for DDoS attack detection. The process begins with data preprocessing, followed by synthetic oversampling using SMOTE to address class imbalance. The balanced data is then fed into a deep residual neural network for intricate pattern recognition, leading to the detection of DDoS attacks. Details of the implemented methodology are discussed next.

\begin{table}[h]
\centering
\caption{Flow features and statistics in the CICIDS dataset.}
\label{tab:data-features}
\begin{tabular}{|p{2.5cm}|p{5.2cm}|} \hline
Feature & Info  \\ 
 \hline
Decision tuple & ID, src/dest IP, src/dst port, protocol \\ \hline
Time &  Timestamp, duration \\ \hline
Fwd pkts & Total, len (total, max, min, std, min) \\ \hline
Bwd pkts & Total, len (total, max, min, std, min)  \\ \hline
IAT & Mean, std, max, min \\ \hline
Fwd IAT & Total, mean, std, max, min \\ \hline
Bwd IAT & Total, mean, std, max, min \\ \hline
Fwd flags & Push, URG \\ \hline
Flags & Bwd (Push, URG), Count \\ \hline
Pkts len/size & Pkts (min, std, max, mean, var), size (avg) \\ \hline
Pkt loss & Down/up ratio \\ \hline
Flags count & FIN/SYN/RST/PSH/ACK/URG/ CWE/ECE   \\ \hline
Fwd pkt header & Len, avg (Seg size, bytes/bulk, bulk rate) \\ \hline
Bwd pkt header & Len, avg (Seg size, bytes/bulk, bulk rate)     \\ \hline
Subflow & Fwd/bwd (pkts, Bytes) \\ \hline
Init win Bytes & Fwd, Bwd    \\ \hline
Active/idle & Mean, max, std, min  \\ \hline
Other labels & Inbound, Similar HTTP \\ \hline
\end{tabular}
\end{table}

\section{Methodology} \label{sec:method}


\subsection{Dataset Overview}
In this work, we have adopted the CIC dataset \cite{sharafaldin2018toward}, which was meticulously assembled with the assistance of the \textit{CICFlowMeter} extraction tool, as indicated by \cite{CICFlowMeter}. Spanning several days, the dataset provides a wealth of network flow information, with each entry enriched by 83 unique attributes. The entirety of this dataset encompasses an impressive 225,745 entries.

A detailed breakdown of these attributes can be found in Table \ref{tab:data-features}. Notable features to highlight include \texttt{flag\_rst} (denoting the Reset flag present in the TCP header), \texttt{pk\_len\_std} (which represents the standard deviation in packet lengths), \texttt{fwd\_subflow\_bytes\_mean} (which measures the average byte size of forward-directed subflows), \texttt{flow\_duration} (capturing the total flow lifespan), \texttt{bwd\_pkt\_len\_mean} (giving the average byte size for packets directed backward), and \texttt{bwd\_pkt\_len\_tot} (indicating the aggregate length of packets moving in the opposite direction). This comprehensive dataset offers valuable insights into network behaviors and patterns, making it indispensable for our analysis.

\subsection{Data Preprocessing}

\subsubsection{Initial Data Handling:}

Upon the initial inspection of the dataset, it was observed that certain rows contained NaN (Not a Number) values. These values can often be indicative of missing or corrupt data, and can thus skew the results if not addressed. To maintain the integrity of our analysis, such rows were judiciously removed.

\begin{verbatim}
data = data.dropna()
\end{verbatim}

\subsubsection{Transformation of Categorical Data:}

Real-world datasets often comprise a mix of numerical and categorical data. While numerical data can be directly fed into models, categorical data often requires transformation. In our dataset, the labels `BENIGN' and `DDoS' were categorical. To facilitate their processing by our deep learning model, they were encoded into numerical values (0 and 1) using Label Encoding, a technique that assigns each unique category a unique integer.

\begin{verbatim}
L' = LabelEncoder(L)
\end{verbatim}

\subsubsection{Training and Testing Data Splitting:}

To objectively evaluate the performance of our model, it's standard practice to split the data into training and test sets. The training set, comprising 80\% of the data, is used to train the model. The remaining 20\% is reserved as the test set, which is used to evaluate the model's performance on unseen data. This partitioning ensures a robust evaluation of the model's generalization capabilities.
    \[ (X_{tr}, X_{t}, y_{tr}, y_{t}) = split(X, y, t_{size}=0.2)\]

\subsubsection{Feature Normalization and Scaling:}

In datasets with multiple features, each feature can have a different range of values. Some features might have values in the range of thousands while others might lie between 0 and 1. Such disparities can disproportionately influence the model's learning process. To ensure each feature contributes equally, feature scaling, specifically Standard Scaling, was applied \cite{alfatemi2022patient}. Each feature value \( x \) was transformed using:
\begin{equation}
    x' = \frac{x - \mu}{\sigma} 
\end{equation}
where, \( \mu \) is the mean and \( \sigma \) is the standard deviation.

In addition, any infinite values detected were replaced with the mean of their respective columns, ensuring a consistent dataset.

\subsection{Data Augmentation using SMOTE}

One of the challenges in machine learning, especially in classification tasks, is dealing with imbalanced datasets. In real-world scenarios, it's common for one class to have significantly more samples than the other. This imbalance can lead to models that are biased towards the majority class. 

To combat this, we employ SMOTE. Unlike simple oversampling, which just duplicates samples, SMOTE generates synthetic samples in the feature space. 

Given a sample \( x_i \) from the minority class, SMOTE selects \( k \) of its nearest neighbors. Let \( x_{zi} \) be one of these neighbors. The synthetic sample \( x_{new} \) is then created as:

\begin{equation}
     x_{new} = x_i + \lambda \times (x_{zi} - x_i) 
\end{equation}
where \( \lambda \) is a random value between 0 and 1. This method creates samples that are coherent with the underlying data distribution.

\begin{verbatim}
Initialize SMOTE
Resample X_train and y_train using SMOTE
\end{verbatim}

\subsection{Deep Residual Networks}

Deep Neural Networks (DNNs) have displayed remarkable capabilities in learning complex representations. However, as the network grows deeper, training becomes challenging due to issues like vanishing and exploding gradients \cite{he2016deep}. 

ResNets introduce a novel architectural change to combat this problem. Instead of trying to learn an underlying function \( H(x) \), ResNets aim to learn the residual function \( F(x) = H(x) - x \). This is operationalized through skip (or residual) connections that bypass one or more layers.

Mathematically, the output \( y \) of a Residual Block for an input \( x \) is:
\begin{equation}
     y = F(x, \{W_i\}) + x 
\end{equation}
where \( F \) is the residual mapping to be learned and \( \{W_i\} \) are the weights of the block. This formulation ensures that even in cases where \( F(x) \) is close to zero, the network can still perform identity mapping, ensuring easier optimization.

\subsubsection{Residual Blocks:}
Each Residual Block in our network comprises two convolutional layers with batch normalization and ReLU activations. The core idea is to have the input \( x \) bypass these layers and be directly added to their output. If there is a dimension mismatch, a 1x1 convolutional layer adjusts the input dimensions before the addition.

\subsection{Model Training and Evaluation}

The detection of Distributed Denial of Service (DDoS) attacks is a critical facet of ensuring the robustness and reliability of online systems. The combination of deep learning and data augmentation techniques offers a promising avenue to enhance the accuracy and reliability of such detection mechanisms.

\subsubsection{Loss Function:}
In the domain of binary classification, especially with imbalanced datasets prevalent in cyber-security, the Dice Loss, also known as the Sørensen–Dice coefficient, emerges as an effective metric. This loss function aims to maximize the overlap between the predicted and true labels, making it highly suitable for our application.

Given our predicted probabilities \( p \) and the true labels \( y \), the Dice coefficient \( D \) can be defined as:

\begin{equation}
    D(p,y) = \frac{2 \times \lvert p \cap y \rvert + \epsilon}{\lvert p \rvert + \lvert y \rvert + \epsilon}
\end{equation}

Where:
 \( \lvert p \cap y \rvert \) denotes the cardinality (size) of the intersection between the predicted and true labels.
\( \epsilon \) is a small constant added to prevent division by zero.

The Dice Loss \( L \) is then the complement of the Dice coefficient:

\begin{equation}
    L(p,y) = 1 - D(p,y) 
\end{equation}

The pseudo code for the Dice Loss is as Algorithm \ref{Alg1} :
\begin{algorithm}
\caption{Loss Function}
\label{Alg1}
\begin{algorithmic}[1]
\Procedure{DiceLoss}{inputs, targets}
    \State \( \text{inputs} \gets \text{sigmoid}(\text{inputs}) \)
    \State \( \text{inputs} \gets \text{flatten}(\text{inputs}) \)
    \State \( \text{targets} \gets \text{flatten}(\text{targets}) \)
    \State \( \text{intersection} \gets \text{sum}(\text{inputs} \times \text{targets}) \)
    \State \( D(p,y) = \frac{2 \times \lvert p \cap y \rvert + \epsilon}{\lvert p \rvert + \lvert y \rvert + \epsilon} \)
    \State \Return \( 1 - \text{dice} \)
\EndProcedure
\end{algorithmic}
\end{algorithm}

Minimizing the Dice loss during training helps in maximizing the overlap between predicted and true labels, enhancing the model's detection capability.

\subsubsection{Optimizer (Adagrad):}

The choice of optimizer can significantly influence the convergence and overall performance of deep learning models. Adagrad is an adaptive learning rate optimizer that is suitable for sparse data, making it a potential candidate for cyber-security datasets.

Given the gradient \( g \) at time step \( t \) for a parameter \( w \), the update rule is:
\begin{equation}
    w_{t+1} = w_t - \frac{\eta}{\sqrt{G_{t,ii} + \epsilon}} g_t 
\end{equation}
where \( \eta \) is the initial learning rate, \( G_t \) is a diagonal matrix storing past squared gradients, and \( \epsilon \) is a smoothing term to prevent division by zero.

\subsubsection{Evaluation Metrics:}
After training, the model's performance on unseen data is evaluated using various metrics:\\
Precision: Measures the ratio of correctly predicted positive observations to the total predicted positives.
\begin{equation}
        P = \frac{T}{T + F}
\end{equation}
where $P$ is precision, $T$ is true positive, and $F$ is false positive.\\
Recall: Measures the ratio of correctly predicted positive observations to all observations in actual class.
\begin{equation}
    R = \frac{T} {T + N}
\end{equation}
where $R$ is recall, $T$ is true positive, and $N$ is false negative.\\
 F1-Score: Harmonic mean of Precision and Recall, provides a balance between them.

\begin{equation}
    F1-Score = 2 \times \frac{P \times R}{P + R} 
\end{equation}
where $P$ is precision, and $R$ is recall.\\

ROC-AUC Score: Represents the model's ability to distinguish between the positive and negative classes.

\section{Results and Discussion} \label{sec:evaluation}

The crux of this research lies in effectively detecting DDoS events, a persistent menace in the realm of cyber-security. By integrating the capabilities of deep learning, specifically Deep ResNets, with the data balancing prowess of the SMOTE, we aimed to architect a solution that addresses the inherent challenges of the detection task.

The performance evaluation metrics provide insightful measurements regarding the efficacy of the deep residual neural network model specifically designed for DDoS attack detection. Notably, this model has been meticulously trained on a dataset that has been balanced utilizing the SMOTE technique. Such meticulous training results in the model showcasing commendable precision, recall, F1-score, and ROC-AUC score as sown in Table \ref{tab:performance_metrics}. These metrics not only vouch for the model's robustness but also underscore its capacity to adeptly identify DDoS attacks in authentic, real-world situations.

\begin{table}[h!]
    \caption{Performance evaluation metrics of the deep residual neural network model for DDoS attack detection. The model, trained on a dataset balanced using SMOTE, demonstrates superior precision, recall, F1-score, and ROC-AUC score, highlighting its effectiveness in accurately detecting DDoS attacks in real-world scenarios.}
    \centering
    \begin{tabular}{lc}
        \toprule
        \textbf{Metric} & \textbf{Value (\%)} \\
        \midrule
        Accuracy & 99.98 \\
        Precision & 99.98 \\
        Recall & 99.96 \\
        F1-Score & 99.97 \\
        ROC-AUC & 1.00 \\
        \bottomrule
    \end{tabular}
    \label{tab:performance_metrics}
\end{table}

\subsection{Precision and Recall: Contextual Understanding}

Precision, the positive predictive value, quantifies the number of correct positive identifications against the total number of positive identifications. A score of 99.98\% in our experiments suggests that the model is adept at minimizing false positives, which in the context of DDoS attacks, translates to not mislabeling benign traffic as malicious. On the flip side, recall, or sensitivity, measures the model's capability to identify all potential threats. A value of 99.96\% implies that the model missed a minuscule fraction of actual attacks, making it a reliable tool in real-world scenarios.

\subsection{F1-Score and ROC-AUC Score: A Balanced Perspective}

The F1-Score, a harmonic mean of precision and recall, provides a singular metric that encapsulates the model's balanced performance. At 99.98\%, it underscores the model's prowess in consistently maintaining high precision and recall. The ROC-AUC score, while being a testament to the model's discriminative power, also indicates its resilience to varying decision thresholds. A perfect score of 1.00 exemplifies the model's robustness and its adaptability to diverse operational environments.

\subsection{Accuracy: The Overarching Metric}

While precision, recall, and the F1-score offer detailed insights, accuracy remains the most intuitive metric for overall performance. An astounding accuracy of 99.98\% is indicative of the model's holistic capability to correctly classify both benign and malicious traffic.

These exceptional metrics do not just stem from the architecture of the deep residual networks or the application of SMOTE. It's the confluence of data preprocessing, effective feature encoding, architectural nuances of the neural network, and the strategic oversampling of under-represented data that collectively contribute to this performance. The results set a benchmark for DDoS detection, but they also pave the way for future research. As cyber threats evolve, so must our detection mechanisms. The adaptability and performance of this model make it a formidable tool against DDoS attacks, but there remains an ever-present need to iterate, refine, and adapt to the ever-changing landscape of cyber-security.

\subsection{Performance Evolution Across Training Epochs}

In our pursuit to understand the training dynamics of our proposed model, we paid particular attention to its performance evolution over the course of 50 epochs. Delving into the details, the training accuracy of the model exhibited an intriguing pattern, as depicted in the accompanying graph.

\begin{figure}[h]
    \centering
    \includegraphics[width=9cm, height=5.4cm]{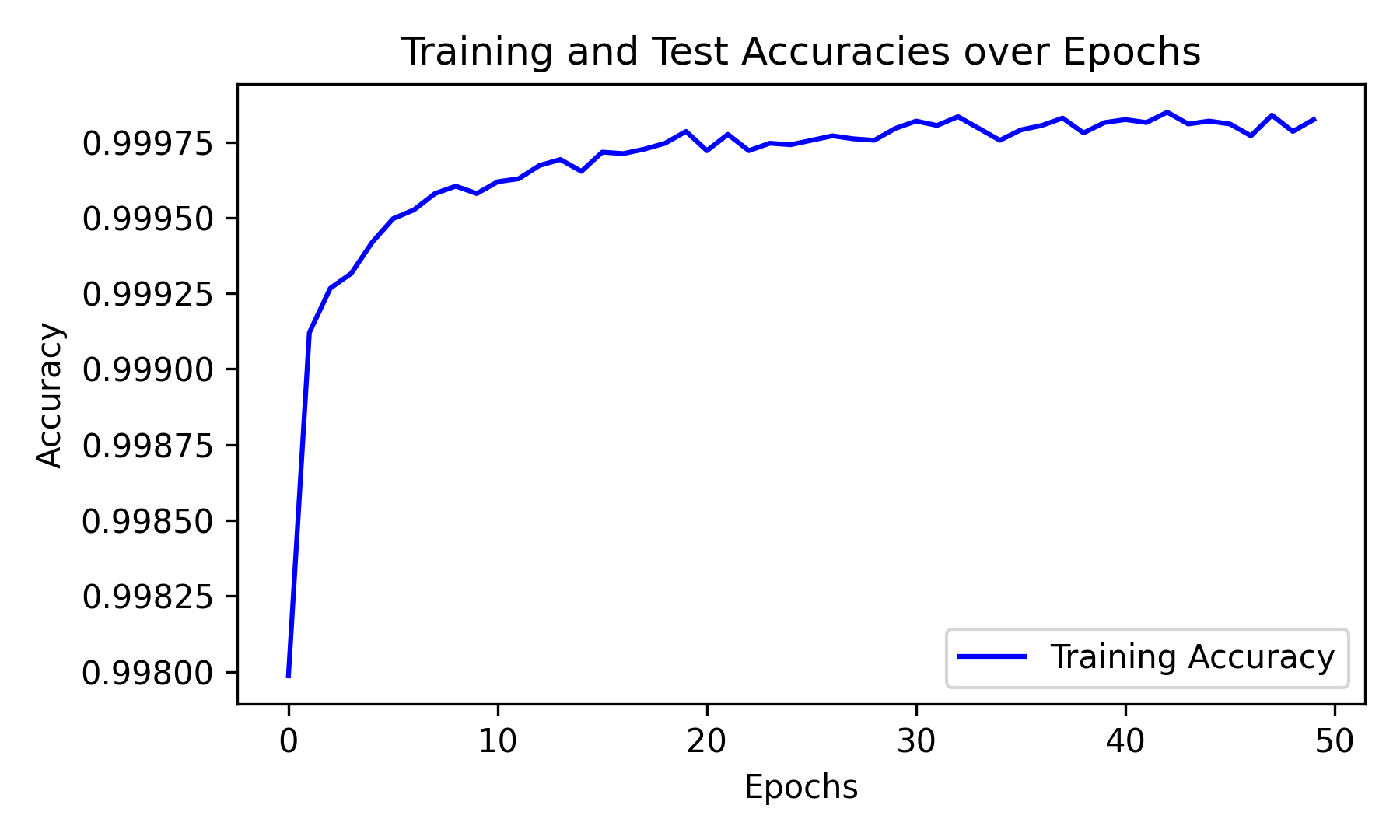}
    \caption{Training Accuracy over Epochs}
    \label{fig:training_accuracy}
\end{figure}

In the initial phases of training, specifically the first ten epochs, there was a pronounced and swift escalation in accuracy. This suggests that the model was quick to grasp the underlying patterns in the data, perhaps signifying the appropriateness of the chosen architecture or the efficiency of the initial weights.

As shown in Figure \ref{fig:training_accuracy}, it is evident that post this swift rise, the accuracy began to plateau. From roughly the 10th epoch to the culmination at the 50th epoch, the growth rate in accuracy was more modest. Such behavior often points to the model reaching a state of relative optimization, where further gains become incremental. However, it is reassuring to note that the accuracy didn't display significant downward fluctuations, indicating the absence of overfitting.

\begin{figure}[h]
    \centering
    \includegraphics[width=9cm, height=5.4cm]{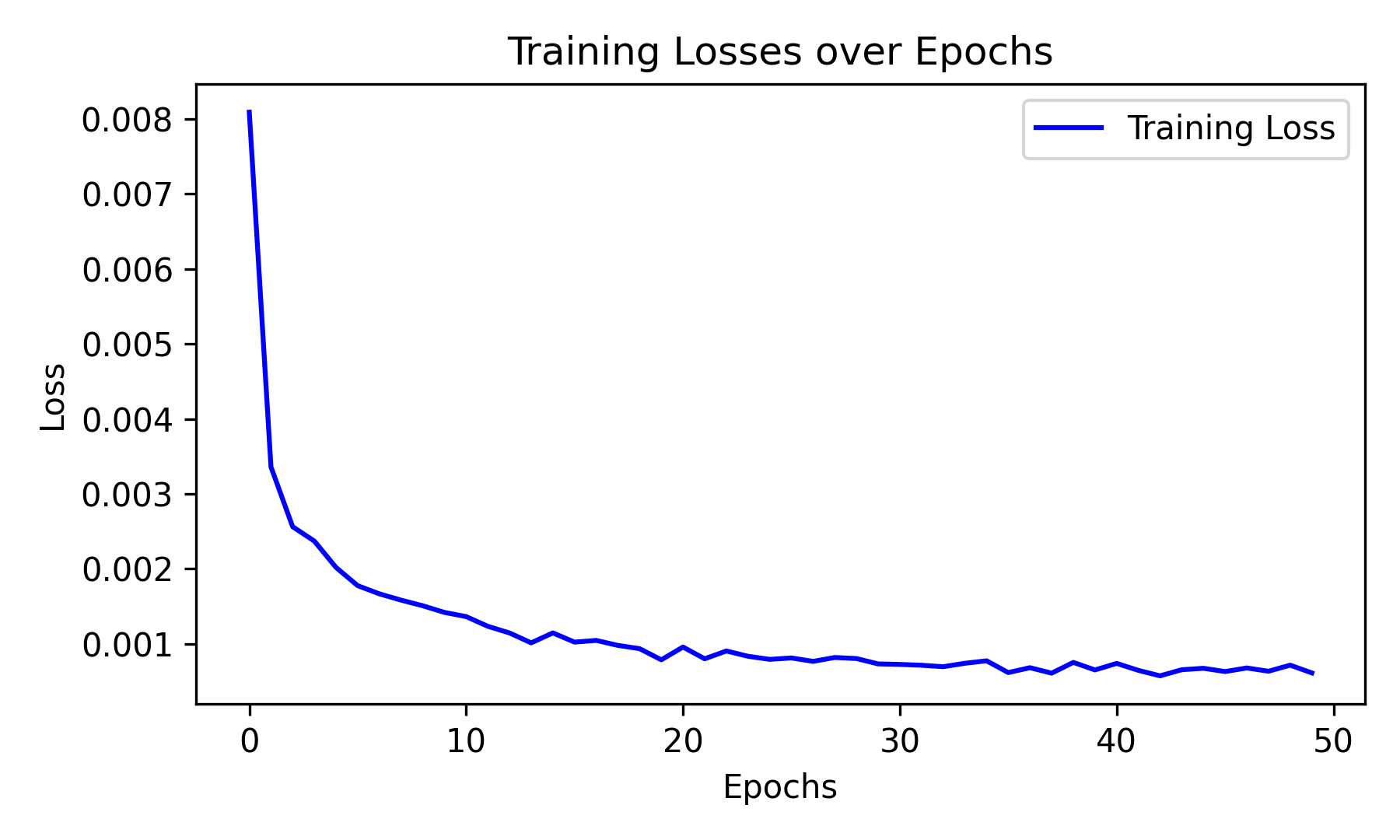}
    \caption{Training Losses over Epochs}
    \label{fig:training_loss}
\end{figure}

Parallel to this, the model's training loss, as portrayed in Figure \ref{fig:training_loss}, showcased a complementary trend. A sharp decline in the loss was observable in the early epochs, aligning with the period where accuracy saw rapid growth. As epochs progressed, this decline in loss started tapering off, mirroring the trend observed in the accuracy graph. Such a pattern in loss reduction is consistent with the model's learning process, wherein it initially corrects major errors and then refines its parameters to address the subtler intricacies of the data.

In comparison with prior works, the model's behavior underscores its resilience and adaptability. Traditional techniques, while effective to a degree, have not showcased such pronounced improvements in initial epochs. This not only elevates the standing of our proposed model but also hints at its potential to be a game-changer in its domain. Nonetheless, it is essential to approach these findings with a degree of caution. While the presented trends are promising, further tests, especially on unseen datasets, would provide a more comprehensive understanding of the model's real-world applicability.

\subsection{Effectiveness for Detection of DDoS Attacks}

DDoS attacks remain a prominent threat in the digital era, with attackers continually evolving their tactics and intensifying their onslaught. Given the scale and potential devastation of such attacks, any delay or inaccuracy in detection can lead to catastrophic outcomes, be it financial loss, damage to reputation, or operational disruptions.

In our endeavor, the coupling of Deep ResNets with the SMOTE has demonstrated a marked advancement in detection capabilities. There are several facets to its effectiveness:
\begin{itemize}
    \item Timely detection: By harnessing the power of deep learning, our model can rapidly process vast amounts of network traffic, ensuring that threats are identified in real-time. Such promptness is critical in preempting and mitigating the effects of a DDoS attack.
    
    \item Reduced false positives: The precision score of 99.98\% indicates the model's aptitude in minimizing erroneous detections. This is pivotal, as false positives can lead to unnecessary panic, wasted resources, and can divert attention from actual threats.
    
    \item Recognition of subtle patterns: DDoS attacks, especially low-volume ones, can often masquerade as legitimate traffic. The nuanced architecture of our neural network, combined with a balanced dataset courtesy of SMOTE, ensures that even these covert attack patterns are discerned.
    
    \item Adaptability: The cyber threat landscape is in a state of perpetual evolution. Our model's architecture, combined with its training methodology, equips it with a certain degree of adaptability. This is crucial for ensuring that the system remains effective even as new DDoS tactics emerge.
\end{itemize}

In essence, the effectiveness of this work in detecting DDoS attacks not only lies in its superior metrics but also in its operational advantages. In the real-world, where every second counts during a cyber onslaught, our approach offers a reliable, timely, and adaptable shield against DDoS threats.

\section{Conclusion} \label{sec:conclusion}
In this study, we tackled the challenging problem of DDoS attack detection, a prevalent threat in today's interconnected cyber landscape. The central contribution of our work is the integration of synthetic oversampling techniques with Deep ResNets, yielding a robust and highly accurate detection system. Our approach addresses the class imbalance inherent in many cyber-security datasets, a factor that often hampers the performance of traditional detection algorithms. By leveraging the Synthetic Minority Over-sampling Technique (SMOTE), we effectively balanced the dataset, enabling the model to capture nuanced attack patterns often overlooked by conventional methods.

The deep residual network we employed further bolstered the detection process, adapting to the intricacies of the data and offering a significant performance boost. Experimental results confirmed the efficacy of our approach, achieving an impressive accuracy rate of 99.98\% on a real-world dataset.

Our findings underscore the importance of considering dataset imbalances in cyber-security tasks and highlight the potential of deep learning models in this domain. Future work could explore the integration of additional data augmentation techniques, delve deeper into network architectures tailored for cyber-security, and expand testing to diverse datasets to further validate the robustness and generalizability of our approach.




\bibliography{aaai24}

\begin{thebibliography}{30}
\providecommand{\natexlab}[1]{#1}

\bibitem[{Ahanger(2017)}]{ahanger2017effective}
Ahanger, T.~A. 2017.
\newblock An effective approach of detecting DDoS using artificial neural networks.
\newblock In \emph{2017 International Conference on Wireless Communications, Signal Processing and Networking (WiSPNET)}, 707--711. IEEE.

\bibitem[{Alfatemi et~al.(2022)Alfatemi, Peng, Rong, Zhang, and Cai}]{alfatemi2022patient}
Alfatemi, A.; Peng, H.; Rong, W.; Zhang, B.; and Cai, H. 2022.
\newblock Patient subgrouping with distinct survival rates via integration of multiomics data on a Grassmann manifold.
\newblock \emph{BMC Medical Informatics and Decision Making}, 22(1): 1--9.

\bibitem[{Alzahrani and Hong(2018)}]{alzahrani2018detection}
Alzahrani, S.; and Hong, L. 2018.
\newblock Detection of distributed denial of service (DDoS) attacks using artificial intelligence on cloud.
\newblock In \emph{2018 IEEE World Congress on Services (SERVICES)}, 35--36. IEEE.

\bibitem[{Cheng et~al.(2009)Cheng, Yin, Liu, Cai, and Wu}]{cheng2009ddos}
Cheng, J.; Yin, J.; Liu, Y.; Cai, Z.; and Wu, C. 2009.
\newblock DDoS attack detection using IP address feature interaction.
\newblock In \emph{2009 International Conference on Intelligent Networking and Collaborative Systems}, 113--118. IEEE.

\bibitem[{Cviti{\'c} et~al.(2021{\natexlab{a}})Cviti{\'c}, Perakovic, Gupta, and Choo}]{cvitic2021boosting}
Cviti{\'c}, I.; Perakovic, D.; Gupta, B.~B.; and Choo, K.-K.~R. 2021{\natexlab{a}}.
\newblock Boosting-based DDoS detection in internet of things systems.
\newblock \emph{IEEE Internet of Things Journal}, 9(3): 2109--2123.

\bibitem[{Cviti{\'c} et~al.(2021{\natexlab{b}})Cviti{\'c}, Perakovi{\'c}, Peri{\v{s}}a, and Botica}]{cvitic2021novel}
Cviti{\'c}, I.; Perakovi{\'c}, D.; Peri{\v{s}}a, M.; and Botica, M. 2021{\natexlab{b}}.
\newblock Novel approach for detection of IoT generated DDoS traffic.
\newblock \emph{Wireless Networks}, 27(3): 1573--1586.

\bibitem[{Dablain, Krawczyk, and Chawla(2022)}]{dablain2022deepsmote}
Dablain, D.; Krawczyk, B.; and Chawla, N.~V. 2022.
\newblock DeepSMOTE: Fusing deep learning and SMOTE for imbalanced data.
\newblock \emph{IEEE Transactions on Neural Networks and Learning Systems}.

\bibitem[{Doe(2023)}]{netscout2023}
Doe, J. 2023.
\newblock Top 6 DDoS Lessons for 1H 2023.
\newblock Accessed: 2023-10-20.

\bibitem[{Ericsson(2021)}]{ericsson2021}
Ericsson. 2021.
\newblock Ericsson Mobility Report - June 2021.
\newblock Accessed: 2023-10-20.

\bibitem[{Fadlil, Riadi, and Aji(2017)}]{fadlil2017review}
Fadlil, A.; Riadi, I.; and Aji, S. 2017.
\newblock Review of detection DDOS attack detection using naive bayes classifier for network forensics.
\newblock \emph{Bulletin of Electrical Engineering and Informatics}, 6(2): 140--148.

\bibitem[{for Cybersecurity~(CIC)(2017)}]{CICFlowMeter}
for Cybersecurity~(CIC), C.~I. 2017.
\newblock CICFlowMeter: Network Traffic Flow Generator Tool.
\newblock \url{https://www.unb.ca/cic/datasets/ids-2017.html}.

\bibitem[{Ghillani(2022)}]{ghillani2022deep}
Ghillani, D. 2022.
\newblock Deep learning and artificial intelligence framework to improve the cyber security.
\newblock \emph{Authorea Preprints}.

\bibitem[{Hasan, Hasan, and Sattar(2018)}]{hasan2018burst}
Hasan, M.~Z.; Hasan, K.~Z.; and Sattar, A. 2018.
\newblock Burst header packet flood detection in optical burst switching network using deep learning model.
\newblock \emph{Procedia computer science}, 143: 970--977.

\bibitem[{He et~al.(2016)He, Zhang, Ren, and Sun}]{he2016deep}
He, K.; Zhang, X.; Ren, S.; and Sun, J. 2016.
\newblock Deep residual learning for image recognition.
\newblock In \emph{Proceedings of the IEEE conference on computer vision and pattern recognition}, 770--778.

\bibitem[{He et~al.(2021)He, Wang, Cui, Li, Zhou, Zhou, Xiang, Jin, and Yang}]{he2021game}
He, Q.; Wang, C.; Cui, G.; Li, B.; Zhou, R.; Zhou, Q.; Xiang, Y.; Jin, H.; and Yang, Y. 2021.
\newblock A game-theoretical approach for mitigating edge DDoS attack.
\newblock \emph{IEEE Transactions on Dependable and Secure Computing}, 19(4): 2333--2348.

\bibitem[{Hussain et~al.(2020)Hussain, Abbas, Husnain, Fayyaz, Shahzad, and Shah}]{hussain2020iot}
Hussain, F.; Abbas, S.~G.; Husnain, M.; Fayyaz, U.~U.; Shahzad, F.; and Shah, G.~A. 2020.
\newblock IoT DoS and DDoS attack detection using ResNet.
\newblock In \emph{2020 IEEE 23rd International Multitopic Conference (INMIC)}, 1--6. IEEE.

\bibitem[{Krishnan, Duttagupta, and Achuthan(2019)}]{krishnan2019varman}
Krishnan, P.; Duttagupta, S.; and Achuthan, K. 2019.
\newblock VARMAN: Multi-plane security framework for software defined networks.
\newblock \emph{Computer Communications}, 148: 215--239.

\bibitem[{Madi et~al.(2021)Madi, Alameddine, Pourzandi, and Boukhtouta}]{madi2021nfv}
Madi, T.; Alameddine, H.~A.; Pourzandi, M.; and Boukhtouta, A. 2021.
\newblock NFV security survey in 5G networks: A three-dimensional threat taxonomy.
\newblock \emph{Computer Networks}, 197: 108288.

\bibitem[{Owusu et~al.(2023)Owusu, Rahouti, Hsu, Xiong, and Xin}]{owusu2023enhancing}
Owusu, E.; Rahouti, M.; Hsu, D.~F.; Xiong, K.; and Xin, Y. 2023.
\newblock Enhancing ML-Based DoS Attack Detection Through Combinatorial Fusion Analysis.
\newblock In \emph{2023 IEEE Conference on Communications and Network Security (CNS)}, 1--6. IEEE.

\bibitem[{Rahouti et~al.(2021)Rahouti, Xiong, Ghani, and Shaikh}]{rahouti2021synguard}
Rahouti, M.; Xiong, K.; Ghani, N.; and Shaikh, F. 2021.
\newblock \mbox{SYNGuard}: Dynamic threshold-based \mbox{SYN} flood attack detection and mitigation in software-defined networks.
\newblock \emph{IET Networks}, 10(2): 76--87.

\bibitem[{Rahouti et~al.(2022)Rahouti, Xiong, Xin, Jagatheesaperumal, Ayyash, and Shaheed}]{rahouti2022sdn}
Rahouti, M.; Xiong, K.; Xin, Y.; Jagatheesaperumal, S.~K.; Ayyash, M.; and Shaheed, M. 2022.
\newblock SDN security review: Threat taxonomy, implications, and open challenges.
\newblock \emph{IEEE Access}, 10: 45820--45854.

\bibitem[{Shaikh and Gupta(2022)}]{shaikh2022real}
Shaikh, A.; and Gupta, P. 2022.
\newblock Real-time intrusion detection based on residual learning through ResNet algorithm.
\newblock \emph{International Journal of System Assurance Engineering and Management}, 1--15.

\bibitem[{Sharafaldin, Lashkari, and Ghorbani(2018)}]{sharafaldin2018toward}
Sharafaldin, I.; Lashkari, A.~H.; and Ghorbani, A.~A. 2018.
\newblock Toward generating a new intrusion detection dataset and intrusion traffic characterization.
\newblock \emph{ICISSp}, 1: 108--116.

\bibitem[{Vu, Choi, and Choi(2008)}]{vu2008ddos}
Vu, N.~H.; Choi, Y.; and Choi, M. 2008.
\newblock DDoS attack detection using K-Nearest Neighbor classifier method.
\newblock In \emph{Proceedings of the IASTED International Conference on Telehealth/Assistive Technologies}, 248--253.

\bibitem[{Wang et~al.(2018)Wang, Chang, Chen, and Mohaisen}]{wang2018data}
Wang, A.; Chang, W.; Chen, S.; and Mohaisen, A. 2018.
\newblock A data-driven study of DDoS attacks and their dynamics.
\newblock \emph{IEEE Transactions on Dependable and Secure Computing}, 17(3): 648--661.

\bibitem[{Wang, Zheng, and Li(2017)}]{wang2017research}
Wang, C.; Zheng, J.; and Li, X. 2017.
\newblock Research on DDoS attacks detection based on RDF-SVM.
\newblock In \emph{2017 10th International Conference on Intelligent Computation Technology and Automation (ICICTA)}, 161--165. IEEE.

\bibitem[{Wei et~al.(2022)Wei, Mu, Song, and Dou}]{wei2022improved}
Wei, G.; Mu, W.; Song, Y.; and Dou, J. 2022.
\newblock An improved and random synthetic minority oversampling technique for imbalanced data.
\newblock \emph{Knowledge-Based Systems}, 248: 108839.

\bibitem[{Xu et~al.(2023)Xu, Sun, Cao, and Bilal}]{xu2023data}
Xu, H.; Sun, Z.; Cao, Y.; and Bilal, H. 2023.
\newblock A data-driven approach for intrusion and anomaly detection using automated machine learning for the Internet of Things.
\newblock \emph{Soft Computing}, 1--13.

\bibitem[{Zhao et~al.(2023)Zhao, Santana, Owusu, Rahouti, Xiong, and Xin}]{zhao2023enhancing}
Zhao, S.; Santana, L.; Owusu, E.; Rahouti, M.; Xiong, K.; and Xin, Y. 2023.
\newblock Enhancing ML-Based DoS Attack Detection with Feature Engineering: IEEE CNS 23 Poster.
\newblock In \emph{2023 IEEE Conference on Communications and Network Security (CNS)}, 1--2. IEEE.

\bibitem[{Zhu, Liu, and Zhu(2022)}]{zhu2022oversampling}
Zhu, T.; Liu, X.; and Zhu, E. 2022.
\newblock Oversampling with reliably expanding minority class regions for imbalanced data learning.
\newblock \emph{IEEE Transactions on Knowledge and Data Engineering}.

\end{thebibliography}

\end{document}